\documentclass[review]{elsarticle}

\journal{ArXiv: 1902.08225}
\usepackage{float}
\usepackage[margin=0.75in]{geometry}
\usepackage{amsmath}
\usepackage{graphicx}
\usepackage[english]{babel}
\usepackage[usenames, dvipsnames]{color}
\usepackage{chngcntr}
\usepackage[utf8]{inputenc}
\usepackage{fancyhdr}
\usepackage{lastpage}
\usepackage{textcomp}
\usepackage{wrapfig}
\usepackage{multicol,caption}
\usepackage{hyperref}
\usepackage{xcolor}

\hypersetup{
    colorlinks,
    linkcolor={red!50!black},
    citecolor={blue!50!black},
    urlcolor={blue!80!black}
}
\setcitestyle{super}

\newenvironment{Figure}
  {\par\medskip\noindent\minipage{\linewidth}}
  {\endminipage\par\medskip}

\begin{document}

	\begin{frontmatter}
		
		\title{Flexible Conductive Composites with Programmed Electrical Anisotropy Using Acoustophoresis}
		\author[a]{Drew S. Melchert}
		\ead{dsmelchert@ucsb.edu}
		\author[b]{Rachel R. Collino}
		\author[c]{Tyler R. Ray}
		\author[a]{Neil D. Dolinski}
		\author[a]{Leanne Friedrich}
		\author[a]{Matthew R. Begley}
		\ead{begley@engr.ucsb.edu}
		\author[a]{Daniel S. Gianola\corref{cor1}}
		\ead{gianola@ucsb.edu}
		\address[a]{Materials Department, Engineering II Building 1355, University of California Santa Barbara, Santa Barbara, CA, 93106, USA}
		\address[b]{MST-7 Engineered Materials, Division of Materials Science and Technology, Los Alamos National Laboratory, Los Alamos, NM 87545 }
		\address[c]{Department of Mechanical Engineering, 2540 Dole St, University of Hawai'i at Manoa, Honolulu, HI 96822}
		\cortext[cor1]{Corresponding author}
\linespread{1.1}
\selectfont
\begin{abstract}
3D printing mechanically flexible composite materials with high electrical conductivity is currently hindered by the need to use high loading of conductive filler, which severely limits flexibility.
Here, microstructural patterning of composite materials via acoustophoresis imparts these materials with high conductivity and flexibility simultaneously, filling a technology gap in the field. Acoustophoresis patterns filler particles into highly efficient percolated networks which utilize up to 97\% of the particles in the composite, whereas the inefficient stochastic networks of conventional dispersed-fiber composites utilize $<5$\%. These patterned materials have conductivity an order of magnitude higher than conventional composites made with the same ink, reaching 48\% the conductivity of bulk silver within the assembled silver-particle networks (at 2.6v\% loading). 
They also have low particle loading so that they're flexible, withstanding $>$500 bending cycles without losses in conductivity and changing conductivity only $5$\% within cycles on average (for 2.6v\% composites). In contrast, conventional unpatterned composites with the same conductivity require such high loading that they're prohibitively brittle.
Finally, modulating the shape of the applied acoustic fields allows control over the anisotropy of the conductive networks and produces materials which are either 2-D conductive, 1-D conductive, or insulating, all using the same nozzle and ink. 
\end{abstract}
\end{frontmatter}
\linespread{1.1}
\selectfont

\section{Introduction}
In flexible conductive composites there is a well-documented trade-off between electrical conductivity and mechanical flexibility that currently limits the viability of 3D printed flexible conductors.\cite{sekitani2010, rogers2010, ray2019} The conductivity of conventional percolated composites has a power law relationship with filler loading, increasing as more particle contacts are formed. The strain tolerance, however, decreases rapidly with the added inclusions in the binder matrix.\cite{sekitani2010,mutiso2015} As a result, highly flexible composites must sacrifice conductivity in conventional approaches.\cite{sekitani2009} This constraint  is particularly important in 3D printed materials which don't have a supporting substrate to absorb stress. As a result, cost-effective flexible electronic devices, much less 3D printed ones, are still out of reach.

Printed conductive composites are typically made with silver or carbon micro- or nano-particles. High conductivity requires maximizing the volume fraction of conductive filler particles, which decreases flexibility so that conventional conductive inks exhibit large (20-600\%) changes in conductivity with bending strain and approximately linear conductivity losses over few cycles, even when supported by a flexible substrate. \cite{ray2019, niu2007, kim2009, merilampi2010} For these 2D printed inks, prestretching the substrate improves flexibility by forming serpentine buckled patterns when relaxed, \cite{ray2019} but this approach is infeasible for freestanding 3D printed structures. High filler loading also poses a problem for printability since higher volume fractions exacerbate clogging and increase viscosity. \cite{deleon2018, postiglione2015, ahn2009} A conventional solution to this problem is to burn off the polymer binder after printing to increase particle contact density, although this step is incompatible with many soft substrate materials. \cite{perelaer2010} While polymer burn-off may not solve the flexibility problem, sintering the conductive particles at higher temperatures improves flexibility as well as conductivity. The temperatures required ($>200^\circ$C), however, are incompatible with most soft materials of choice. \cite{ahn2009} Alternative approaches like laser or microwave sintering may be compatible with 3D printing soft materials, although this approach would introduce complex post-printing steps to an already costly and complex process of printing conductive and support inks with multiple nozzles. \cite{perelaer2006,truby2016} As another approach, liquid eutectic alloy-infiltrated channels in soft materials have a promising combination of stretchability and conductivity, but $>0^\circ$ C melting temperatures, cost, and safety concerns (e.g. Ga leaching) need further development. \cite{dickey2017} Outside of printable materials, soft-substrate lithographic techniques are successfully tackling flexible chips\cite{ray2019,saengchairat2017} but are infeasible for large-area electrodes or device-scale interconnects. \cite{sekitani2010} Thus there is a technology gap in 3D printing soft components with integrated electrical circuits, which necessitates expensive and time-consuming wiring steps. \cite{ahn2009,macdonald2016} Filling this technology gap would substantially increase the cost effectiveness and design complexity of 3D printed soft electronic components, a major boon for applications like soft robotics \cite{rus2015} and wearable electronics. \cite{valentine2017}

\begin{figure}[ht]
	\centering\includegraphics[width=\linewidth]{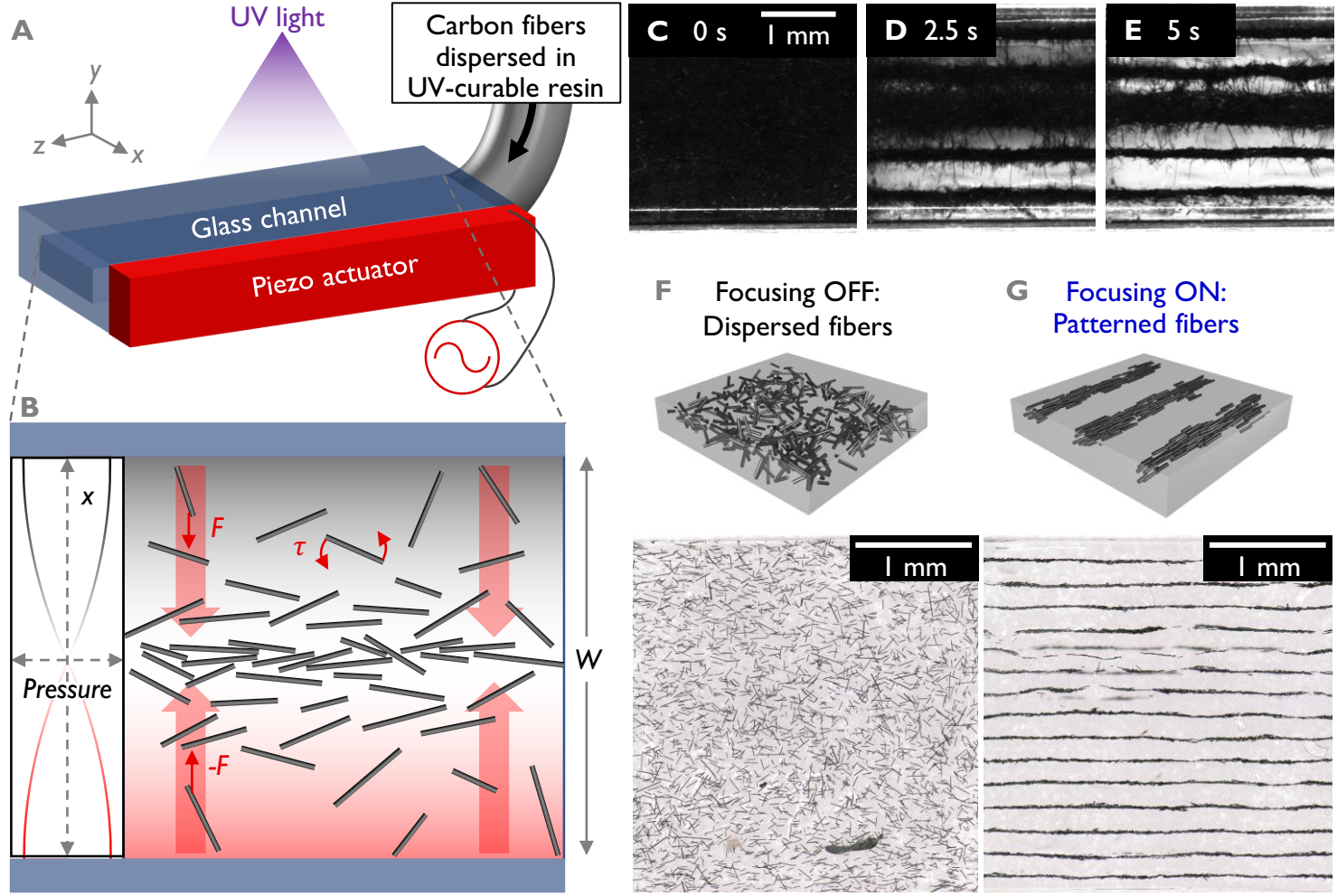}
	\caption{\textbf{A} Schematic of acoustic focusing device. \textbf{B} Diagram of forces aligning and pushing fibers to a low pressure node in a standing pressure half-wave. \textbf{C,D,E} Time-lapse of fibers patterned into parallel bundles by acoustic focusing in photopolymer resin, reaching equilibrium positions after 5-6 seconds. \textbf{F} Illustration and micrograph of an unpatterned composite with carbon fibers dispersed in acrylate resin (0.36v\% carbon fiber). \textbf{G} Illustration and micrograph of a patterned carbon fiber composite (0.36v\% carbon fiber, the same ink as in \textbf{F}) fabricated with acoustic focusing.}
	\label{fig1}
\end{figure}

Here we present an approach for bypassing the trade-off between conductivity and flexibility in composites by assembling filler particles into percolated networks within a flexible polymer matrix. We employ acoustophoresis to compact filler fibers into bundles instead of the stochastic networks formed in traditional dispersed-fiber composites. We demonstrate that these efficiently patterned materials have an order of magnitude higher conductivity than the dispersed-fiber composites made with the same ink and are conductive at an order of magnitude lower particle loading. These patterned materials are also flexible, withstanding at least 500 cycles without losses in conductivity and changing very little with bending strain. Finally, we demonstrate control over transport anisotropy so that 2D conductive, 1D conductive, and insulating materials can be made all using the same ink and same nozzle, simply by modulating acoustic assembly parameters.

\begin{Figure}
	\centering\includegraphics[width=.5\linewidth]{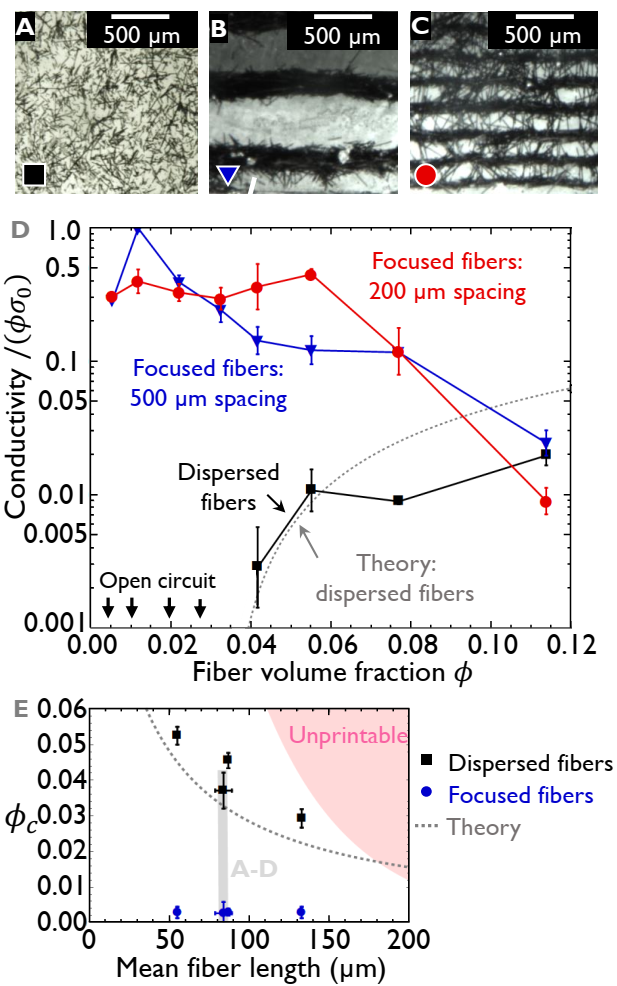}
	\captionof{figure}{ \textbf{A} Micrograph of a 0.5v\% composite made without acoustic focusing. \textbf{B} Micrographs of 2.6v\% composites made with acoustic focusing with 500 \textmu m focused bundle spacing and \textbf{C} 200 \textmu m spacing. \textbf{D} Electrical conductivity normalized by the conductivity of the fibers $\sigma_0$ and the volume fraction $\phi$, which gives the volume fraction of fibers in the composite that contribute to the conductive network. Error bars represent standard deviation, and the dashed line represents a theoretical model. Fiber alignment by shear flow, a factor especially apparent in the higher-viscosity inks with high $\phi$ is expected to decrease conductivity below this model. \cite{mutiso2015} \textbf{E} The percolation threshold $\phi_c$ at which electrical behavior of composites transitions from insulating to conductive. Fiber alignment is expected to increase $\phi_c$, and high spread in the fiber length distribution is expected to decrease $\phi_c$. \cite{mutiso2015} Composites in \textbf{A-D} have the fiber length distributions with high standard deviation highlighted in grey in \textbf{G}. The “unprintable” region indicates approximately which volume fractions and fiber lengths clog the devices.}
	\label{fig2}
\end{Figure}

Acoustophoresis, also called acoustic focusing, is a field-assisted assembly technique that employs pressure fields to rotate and translate particles within a fluid. Acoustophoresis allows for particle positioning, in contrast to electromagnetic field assisted assembly \cite{kokkinis2015,Martin2015,loth2012} or hydrodynamic shear alignment\cite{compton2014,tekinalp2014,lewicki2017} which can only rotate particles.
Acoustic focusing can also handle a larger variety of particle and fluid materials than electromagnetic assembly because it depends only on the contrast in density and compressibility between the particle and fluid material, not polarizability or conductivity or other material properties. Other stochastic approaches like freeze-casting or laser-directed alignment \cite{xiong2016} show promise for improving material properties of composites, but these are generally restricted to specific material systems and microstructural configurations. Acoustophoresis is highly also scalable, assembling \textmu m- to mm-size structures in seconds, \cite{collino2018,greenhall2014,reyes2018} in contrast to self-assembly which is limited by diffusion rates and is prohibitively slow to assemble $>$10 \textmu m structures. Though used primarily for cell sorting and other biological characterization, \cite{hagsater2007} acoustic focusing has recently been investigated for use in 3D printing via direct ink writing \cite{collino2018, collino2015,collino2016} and stereolithography, \cite{greenhall2017,yunus2017} potentially with user-defined particle patterns. \cite{greenhall2017, greenhall2016, melde2016} Electrical elements have been printed with the latter method, \cite{greenhall2017, yunus2017} although the acoustic reservoir print bath used in stereolithography limits it to producing $<50$ mm components; our direct ink writing method is scalable to larger-scale applications. We investigate assembly and printing dynamics of printing with acoustophoresis elsewhere \cite{collino2018, collino2015, collino2016, friedrich2017} (a discussion of printing rates and resolution can be found there) as an approach which avoids the filler particle size and resolution limitations, material compatibility issues, and the high production time, complexity, and cost associated with multiple-nozzle printing of embedded functional elements. \cite{truby2016, derby2010,shofner2003,ning2015,tekinalp2014}

Acoustophoresis utilizes pressure fields to manipulate particles suspended in a fluid---here, ink in a printing nozzle. Pressure fields are generated by resonating the channel walls of the glass nozzle with a piezoelectric actuator, \cite{king1934,20000152384,bruus2012} as shown in Figure \ref{fig1}A. When the walls are oscillated at the channel's fundamental resonant frequency, $f_1=c/(2W)$ where $c$ is the speed of sound in the ink and $W$ is the channel width, a sinusoidal pressure field $P_1\propto\sin(2\pi x/W)$ is established across the channel. Particles are pushed to the nearest of $n$ low-pressure nodes spaced equally across the nozzle, as shown in Figure \ref{fig1}, and settle due to gravity. A thorough treatment of the acoustic forces and torques on rod-like particles is given in the Supplementary Information \S 1.

To demonstrate spatial programming of electrical conductivity using acoustophoresis, we employed a two-component ink consisting of short carbon fibers or silver-coated glass fibers in acrylate photopolymers. First, composites were UV-cured without acoustic focusing, with fibers uniformly dispersed, as a control case representing conventional 3D-printed composites. Next, using an identical precursor composition, the fibers were focused into bundles using acoustophoresis before the composite was UV-cured, yielding patterned composites. We will discuss the 1D electrical properties of rigid composites made with either carbon fibers for their consistency in conductivity or silver-coated glass fibers for their high conductivity first, before moving onto network anisotropy and flexibility.

\section{Results and Discussion}
\subsection{Dispersed-fiber composites follow the predictions of percolation theory}

The control-case dispersed-fiber composites, made without acoustic focusing, are shown in Figure \ref{fig2}A. These unpatterned composites are insulating at fiber volume fractions $\phi<3.3$\%, resulting in an open-circuit measurement (Figure \ref{fig2}D). Above 3.3\%, these composites have low conductivity ($<1$ S/m), with around 1\% of the fibers contributing to the conductive network. This transition from insulating to conductive at $\phi=3.3\%$ is termed the percolation threshold $\phi_c$, given by $\phi_c=1/(\pi/2a+2a+3+\pi)=3.3\%$, where $a=l/d$ is the aspect ratio of the fibers (assuming monodispersity and random fiber orientation). \cite{mutiso2015}
Below this threshold, too few particles are in contact to form a continuous network, so that charge transport across the material is blocked by regions of insulating matrix material. Above $\phi_c$, however, a continuous conductive network is formed across the material. Percolation theory models the conductivity of two-phase composite materials like these as $\sigma=\sigma_0 (\phi-\phi_c )^2$ for $\phi>\phi_c$, \cite{mutiso2015} where $\sigma_0$ is a fitting parameter that gives the particle and contact resistance, \cite{foygel2005} and the exponent value of 2 describes a 3D network of low aspect-ratio ($a<100$) fibers. Conductivity is expected to be overestimated by this model due to hydrodynamic alignment of fibers during loading of the channel, \cite{mutiso2015} as is indeed observed in our data at higher loading where higher viscosity increases the effect of shear flow particle alignment. The particle and contact resistance  $\sigma_0$ for carbon fibers is 1000 S/m, two orders of magnitude lower than the nominal conductivity of the fibers alone, indicating that there is significant contact resistance between the carbon fibers (as expected).

\subsection{Acoustically patterned composites have high conductivity at low fiber loading}

Patterned composites formed with acoustophoresis, on the other hand, shown in Figure \ref{fig2}B,C have dramatically higher conductivity than the unpatterned composites, as shown in Figure \ref{fig2}D. This is due to efficient use of particles, forcing fillers into percolated bundles instead of relying on random contacts in a stochastic network. Below the percolation threshold for dispersed-fiber composites $\phi_c$, where the unpatterned composites are insulating, the patterned composites have high conductivity. Up to 97\% of the fibers by volume contribute to the conductive network at $\phi=1.2\%$ because fibers preferentially stack end-on-end due to scattering forces in the $x$-direction, \cite{collino2015} a configuration which optimizes filler utilization for 1D transport. Thus as opposed to needing a high concentration of fibers to form robust percolating networks in dispersed-fiber materials, forcing fibers into efficient configurations allows high conductivity even at low loading. At volume fractions up to 8\%, the conductivity of the acoustically patterned composites is at least an order of magnitude higher than that of unpatterned percolated composites made with the same precursor ink due to efficient particle distribution.

\begin{figure}[ht]
	\centering\includegraphics[width=1.\linewidth]{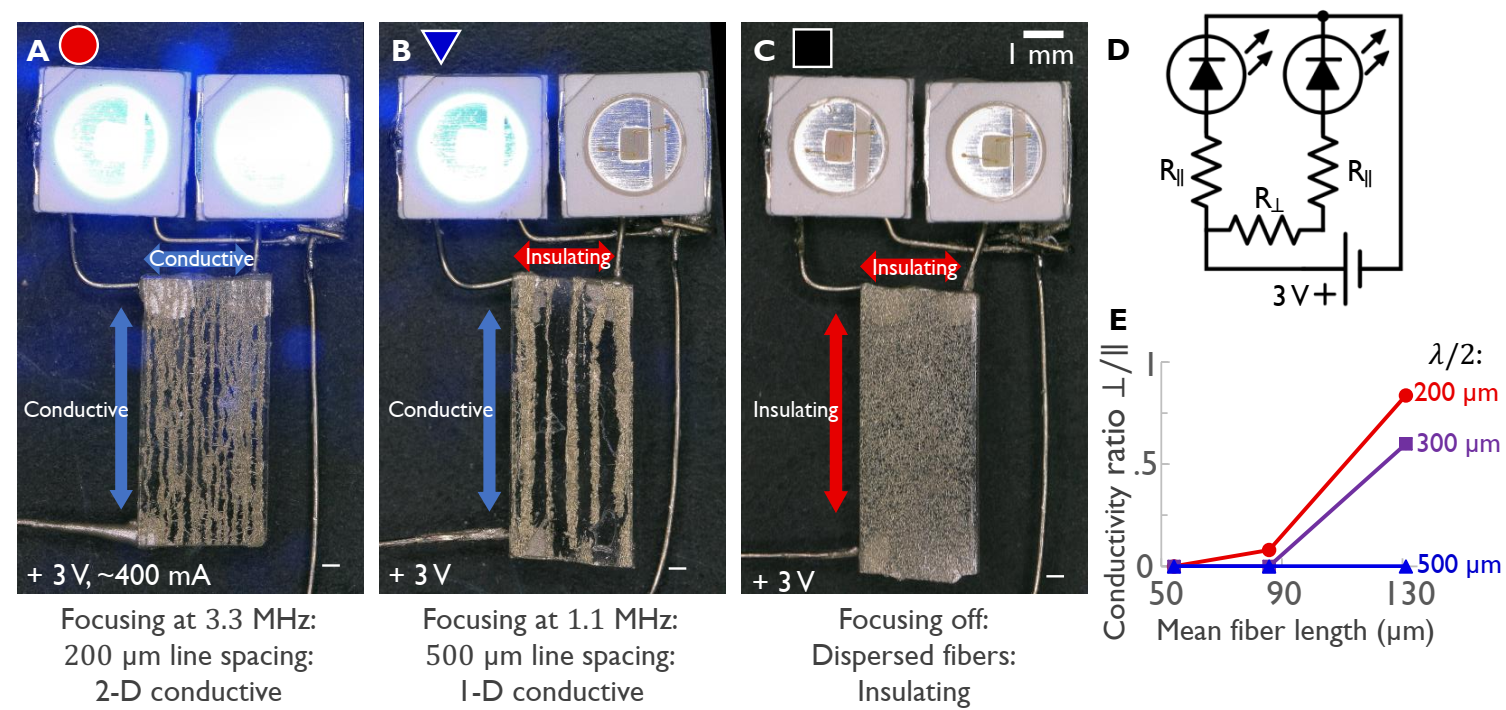}
	\caption{Three composites made using the same precursor ink, 2.6\% silver-coated glass fiber in acrylate resin: \textbf{A} An isotropic conductive composite with focused bundle spacing $\sim200$ \textmu m completing a 3 V circuit to illuminate two 100 mA LEDs. Differences in LED brightness are due to tilt of the metallic reflector with respect to the camera. \textbf{B} An anisotropic conductive composite with bundle spacing $\sim500$ \textmu m completing a 3 V circuit to illuminate only one LED. \textbf{C} An insulating composite made with acoustic focusing turned off. \textbf{D} Schematic of the circuit in \textbf{A-C}. \textbf{E} Dependence of transport anisotropy in 2.6\% composites on mean fiber length and bundle spacing, given as the ratio of conductivity measured in the direction perpendicular to the focused lines to that measured parallel.}
	\label{fig5}
\end{figure}

Within the focused bundles themselves this high conductivity is due to high density of contacts between fibers, as revealed by cross-sectional SEM images and X-ray CT scans of focused fiber bundles, as shown in Figure \ref{fig3}. The (non-normalized) conductivity of composites  is roughly constant at 10 S/m for 1-10\% loading, and is Ohmic (Supplementary Information Figure 2), compared to $<1$ S/m for unpatterned composites. This is aproximately the same conductivity achieved in printable bundled-CNT (carbon nanotube) composites, \cite{xiong2016} despite that CNTs have orders of magnitude higher conductivity than our carbon fibers. Since acoustophoresis can handle metallic filler materials, further enhanced conductivity is possible via higher conductivity fillers. For silver-coated glass fibers (Figure \ref{fig5}), conductivity is $>$5,000 S/m, calculated using the entire volume of 2.6\% composites. These values are an order of magnitude higher than those reached in dispersed-fiber composites even at $\phi=12\%$, the maximum loading before the nozzle clogs during loading, even though the particle patterns are not always ideally straight (improvements in device design are expected to further increase conductivity values).

At the lowest loading value investigated, 0.4\%, focused lines break continuity due to particle scarcity, so that conductivity drops. At volume fractions above 1.2\%, the contact efficiency decreases somewhat as some fibers are oriented radially outward from the bundles (not contributing to the network) but remains an order of magnitude higher than the unpatterned composites until 12\%. At 12\%, the maximum printable loading before channels clog, the normalized conductivity decreases to 1-5\% because such high solid content produces scattered-wave interactions \cite{greenhall2017} and distorted pressure fields which disrupt assembly. In contrast to the strong scaling in dispersed-fiber composites, conductivity in acoustically patterned composites is invariant from 1-10\%. This insensitivity to loading provides freedom to orthogonally control other material properties like stiffness, strength, or thermal conductivity by tuning the filler loading while maintaining high electrical conductivity.

\begin{Figure}
	\centering\includegraphics[width=.5\linewidth]{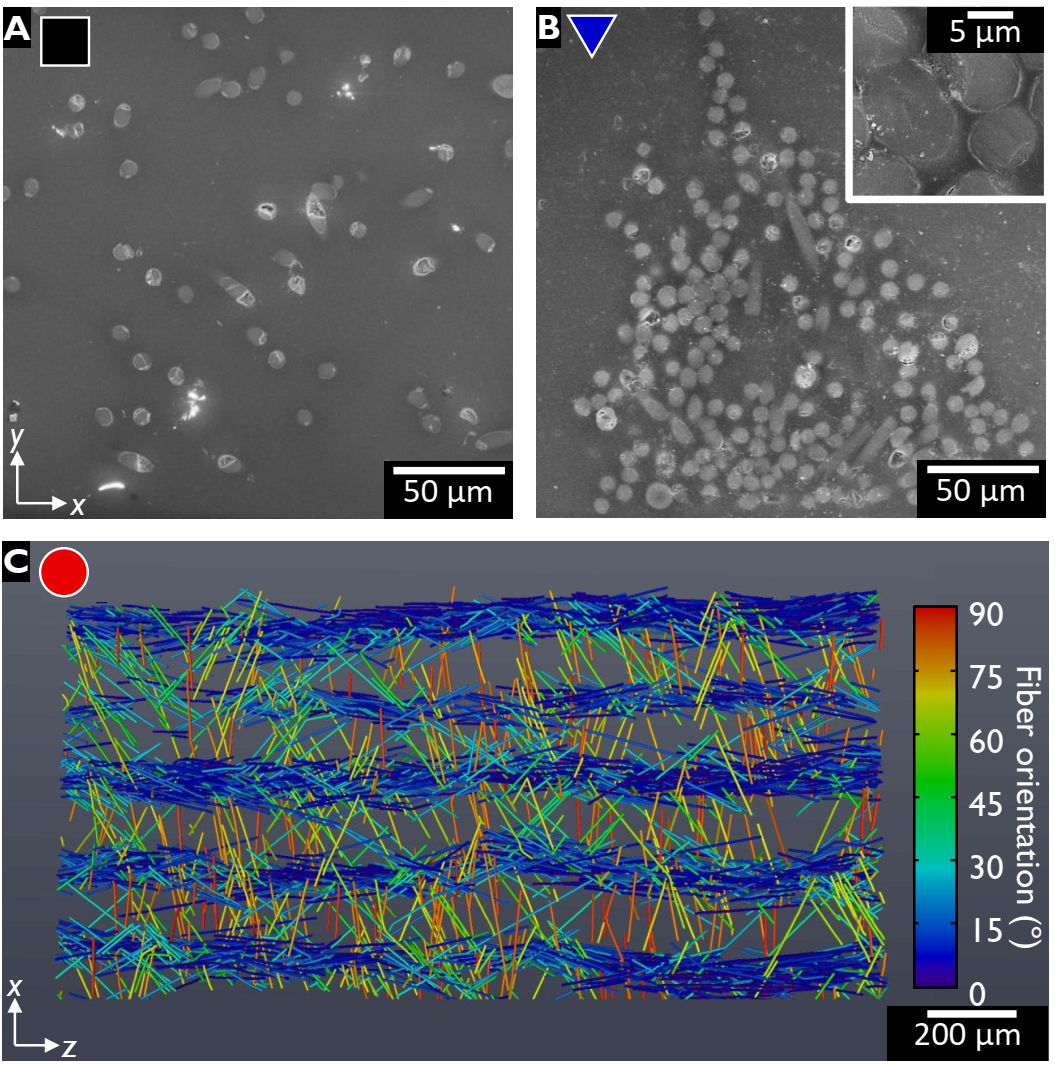}
	\captionof{figure}{SEM images of cross-sections of 2.6\% carbon-fiber composites with \textbf{A} no focusing (insulating) and \textbf{B} ~500 \textmu m focused bundle spacing (anisotropic conductive). \textbf{C} X-ray computed tomography image with cylinders fit to fibers of a 2.6\% carbon-fiber composite with ~200 \textmu m focused line spacing (conductive), with fibers colored by their angle with the z axis.}
	\label{fig3}
\end{Figure}

Additionally, the critical volume fraction for conductivity in patterned composites is an order of magnitude lower at 0.36\% than the dispersed-fiber composites, as shown in Figure \ref{fig2}E. For unpatterned composites the percolation threshold is strongly dependent on fiber length, so that long fibers are better for conductivity (but worse for clogging).\cite{mutiso2015} For patterned composites, the critical loading for conductivity is independent of fiber length because fibers are stacked end-on-end, although conductivity decreases for short fibers because of the added contact resistance. This further adds opportunities to control mechanical or other properties via fiber length while maintaining conductivity.

\subsection{Electrical anisotropy is controlled by focusing wavelength}

The conductivity reported above is one dimensional, running parallel to the focused bundles of carbon fibers, where bundles are insulated from each other. Control over conductivity in the perpendicular direction, however, is possible via changing properties of the applied pressure field. Modulating the focusing wavelength controls the bundle spacing $\lambda/2$, which in turn controls the number of fibers bridging between bundles: fibers longer than $\lambda/4$ are acoustically stable perpendicular to the focused bundles, bridging between bundles. \cite{yamahira2000} This in turn controls the density of transport pathways between bundles in the perpendicular direction and anisotropy in conductivity in the material. The focused bundle spacing is equal to the half-wavelength of the pressure field: focusing at $f=1.1$ MHz yields spacing $\lambda/2=200$ \textmu m; focusing at $f=3.3$ MHz yields $\lambda/2=500$ \textmu m, and so on (Supplementary Information Figure 3). As shown in Figure \ref{fig5}E, composites made with long fibers and close bundle spacing (mean length 130 \textmu m, $\lambda/2=200$ \textmu m, shown in Figure \ref{fig3}C) have a high density of bridging fibers and, as a result, nearly-isotropic conductivity. Conductivity measured in the direction perpendicular to the focused lines (denoted $\perp$) is 84\% that measured in the parallel direction ($\parallel$). At the widest spacing, 500  \textmu m, even using the longest fibers bundles are insulated from each other in the perpendicular direction and the material is totally anisotropic. Acoustics theory models of fiber bridging behavior agree with our data (Supplementary Information Figure 1). By controlling the focused line spacing, the electric conductivity of the patterned composites can be modulated between anisotropic or nearly isotropic. This could be accomplished on-the-fly in a printing modality to allow programmable spatial control of electrical interconnects embedded in printed components.


\subsection{Patterning silver-based and elastomeric conductive composites}

In terms of non-normalized conductivity, silver-coated glass fibers impart conductivity of $>$5000 S/m to patterned composites, or 48\% of the conductivity of bulk silver normalizing to the volume of silver in the composites, which indicates a low contact resistance and high particle utilization. This normalized value compares well to conventional inks, which typically only reach values $>$30\% after polymer burn-off and particle sintering steps ($>200^\circ$ C). \cite{perelaer2010} The non-normalized conductivity is lower because of the low volume of silver in the silver-coated fibers, but is nonetheless on the same order of magnitude as the highest-conductivity long-CNT composites with $\phi>$10wt\%. \cite{bauhofer2009} These materials have ample conductance for mm-scale soft-robotics applications, which typically require $<$100 mA for 1 N actuators. \cite{levard2012,cortes2003} As a demonstration, a 2.6v\% composite completes a circuit to supply $\sim$400 mA to 3 V LEDs in Figure \ref{fig5}. Anisotropy is modulated by choosing focusing frequencies that either result in close spacing and fiber bridging, and therefore nearly isotropic conductivity as in Figure \ref{fig5}A, or wide spacing and therefore anisotropic conductivity as in Figure \ref{fig5}B. The unpatterned composites made with the same precursor ink but without acoustic focusing, Figure \ref{fig5}C, are sub-percolation and thus insulating. Thus, during printing one can define where conductive or insulating material is patterned all while printing using a single nozzle and ink to define simple circuits and interconnects in components.

In composites formed with elastomeric matrix material (1:1 methyl:cyclo hexyl acrylate + 2wt\% butanediol diacrylate, glass transition temperature $14.5^\circ$C), conductivity values in the undeformed state are unchanged, but the composites are flexible as shown in Figure \ref{fig4}. In a bending test, cycling between bending strains of $\epsilon=22\%$ (mandrel radius 0.7 mm, with the particle pattern on the outer face so that it's in tension) and $\epsilon=0\%$ at a rate of 0.5 Hz yielded nearly constant conductivity over 500 cycles, with conductivity recovering to $\sim$90\% of the pristine value (34 S/m) in each cycle. Additionally, conductivity changes only 5\% on average within cycles. This robustness to strain is due to the confinement of jammed, stiff conductive particles joined by freely rotating and sliding contacts within a compliant, stress-absorbing matrix. In contrast, conventional silver conductive inks which rely on silver or graphene micro-flakes exhibit large (20-600\%) changes in conductivity with bending strain and roughly linear conductivity loss with cycle number. \cite{niu2007,kim2009,merilampi2010} As shown in Figure \ref{fig4}E, because conventional dispersed-fiber composites require much higher loading for conductivity, they have both lower conductivity and flexibility, cracking at low strains. Elastomer formulations with lower glass transition temperature or lower cross-linking density resulted in slower recovery times and lower cyclic durability (Supplementary Information Figure 4).

\begin{Figure}
	\centering\includegraphics[width=\linewidth]{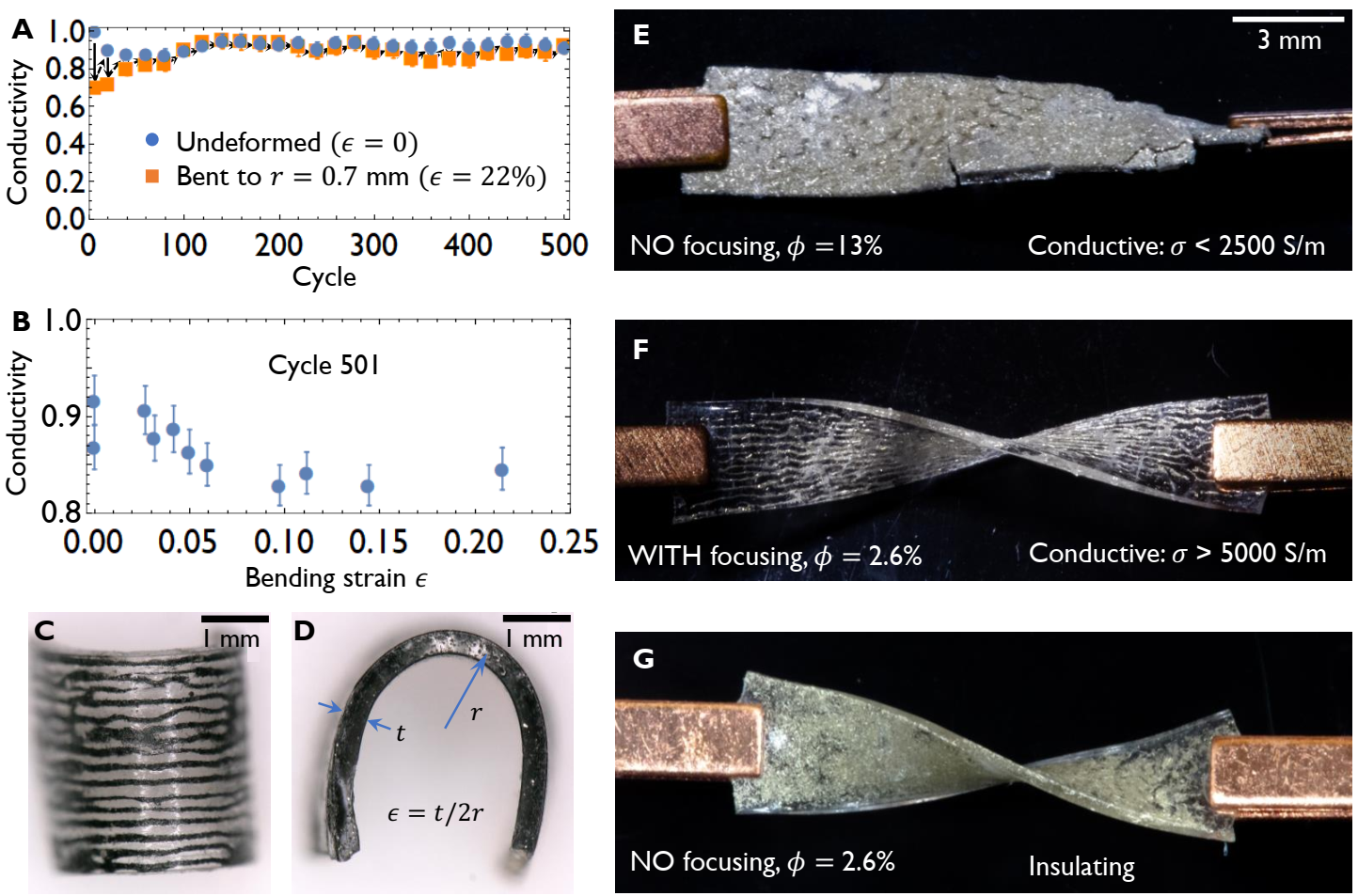}
	\captionof{figure}{\textbf{A} Conductivity of elastomeric patterned composites after bending cycles to $\epsilon=t/2r=22\%$, where $t=300$ \textmu m is the composite thickness, as a fraction of the pristine conductivity (34 S/m). \textbf{B} Conductivity within a single cycle (\#501). \textbf{C,D} Micrographs of a 2.6\% carbon-fiber elastomeric composite. \textbf{E,F,G} Photographs of silver-coated fiber elastomeric composites, twisted to showcase the gains in flexibility and conductivity via acoustic patterning. In \textbf{E}, the left edge of the composite is a fractured edge, having failed completely at a 90 degree twist before being photographed.}
	\label{fig4}
\end{Figure}

\section{Conclusions}
In summary, patterning filler particles into highly efficient percolated networks in composites increases their conductivity by an order of magnitude over conventional dispersed-fiber composites made with the same ink, due to much higher particle utilization efficiency in the percolated networks. Conductivity reaches 48\% that of bulk silver within the conductive networks, a value that typically requires sintering not compatible with desirable soft materials. The composites require an order of magnitude lower particle loading to be conductive compared to dispersed-fiber composites. As a result, the patterned composites withstand $>$500 bending cycles to a radius of 0.7 mm without losses in conductivity, with very low change in conductivity within cycles. This technique is compatible with 3D printing of a wide variety of particle and soft matrix materials, whereas the current alternatives for optimizing conductivity are not.
Finally, we demonstrate that modulating assembly parameters allows patterning of either 2-D conductive, 1-D conductive, or insulating material, all using the same nozzle and ink. This technique is a novel approach to manipulating microstructure to enhance and modulate material properties during printing, paving the way for printing soft components with embedded electrical interconnects or other functional elements.
\footnotesize

\section{Experimental Section}
\subsection{Materials and Particle Suspension}
Milled carbon fibers (Nippon Granoc XN-100-15M; 7$\pm$1 \textmu m in diameter and 80$\pm$60 \textmu m long) were filtered with sieves into populations with length distributions of 60$\pm$20 \textmu m, 90$\pm$40 \textmu m, and 130$\pm$70 \textmu m (Figure \ref{fig2}). The UV-cure monomer consisted 1,6-hexanediol diacrylate (HDDA) (80\% with 100ppm quinone inhibitor, Sigma Aldrich) and 2.5wt\% photoinitiator (1-hydroxycyclohexyl phenyl ketone, Sigma Aldrich). The fibers were weighed dry with a microbalance and resin volumes were drawn with micropipettes, then vortexed and loaded into focusing channels via syringes.

Elastomeric composites are made using 1:1 methyl:cyclo-hexyl methacrylate monomers (both Sigma Aldrich, $>$97\%) with 2\% butanediol diacrylate crosslinker (Sigma Aldrich, 90\%) and 2.5wt\% photoinitiator. Silver-coated fiber composites are made using silver-coated glass rods with 150 \textmu m nominal mean length and 10 \textmu m nominal diameter (Potters Conduct-o-fil AGCLAD12) and HDDA.
\subsection{Acoustic Focusing Device}
Focusing channels were constructed from borosilicate capillaries, 0.3 $\times$ 3 mm rectangular cavity cross section (Borotubing Vitrotubes) cut to 15 mm and attached to metal ferrules (0.25 mm$^2$ cross-section, McMaster) with polyolefin heat-shrink and Epoxy (5-min, Devcon) for loading with syringes via plastic tubing (TYGON, .04” ID). The channels were fit into a focusing device, which consists of a piezoelectric transducer superglued to a machined aluminum channel holder (1 mm wall thickness). The piezoelectric transducers (2 mm thick PZT26, 1 MHz thickness mode, Steminc SMPL20W15T21R111) were soldered to BNC connectors and driven with a signal generator, power supply and amplifier (Fluke 294, Tenma 72-7245, Minicircuits LZY-22+) and monitored in parallel with an oscilloscope (Agilent). The channel holder featured a machined lip to hold the glass capillaries snugly in place and was acoustically coupled to the glass capillary with ultrasonic coupling gel (PosiTector). Capillaries were clamped gently in place with rubber bands to avoid compressing the piezo. An air gap was machined in the holder below the capillaries to block pressure wave transmission in $y$- and $z$-directions and thereby increase pressure field uniformity. The piezo was fixed to the side of the channel holder so that the actuation direction was oriented parallel to the focusing direction in the channel. \cite{iranmanesh2013,muller2012} The piezo was coupled to a water-cooled copper heatsink with thermal couplant (Wakefield Type 120) to hold the piezo temperature, measured with a thermocouple, at ambient temperature.

\subsection{Patterning and Solidification}
The unpatterned dispersed-fiber samples made without acoustic focusing were UV-cured immediately in the device (in 3-5 seconds after dispersion, to avoid letting fibers settle) with a 365 nm UV LED (EXFO Omnicure LX400 with focusing lens, $\sim$1 W/cm$^2$) for one minute ($<$0.5 seconds to gel, 1 second to solidify).

Patterned samples made with acoustic focusing were made by driving the piezoelectric actuator at a resonant frequency of the piezo-channel system. Resonant frequencies were identified as sharp valleys in the impedance magnitude spectrum (Supplementary Information Figure 3). At 3.3\% carbon fibers in HDDA by volume, resonances were measured at 0.660MHz, 0.948MHz, 1.104 MHz, 3.292 MHz, which are the fourth, fifth, sixth, and fifteenth harmonics of the fundamental resonant frequency (confirmed by counting the number of particle lines formed and by fitting to $f_n=nc/2W$). For other particle volume fractions, the resonant frequencies change slightly due to changes in the speed of sound through the resin-fiber suspension. \cite{han2017} These resonances were found manually by observing particle patterns through a stereoscope and adjusting the driving frequency until particle patterns were straight and uniform. The piezo actuator center driving frequency (dictated by piezo thickness) was chosen to match these observed device resonant frequencies: 0.667, 1.0, and 3.0 MHz piezo resonances overlap the channel resonances and result in strong excitation. 
The piezoelectric transducer was driven at 20 Vpp, so that particles reached their equilibrium patterns within six seconds. Higher driving voltages form patterns more quickly ($V\propto\sqrt{E_0}$; \cite{barnkob2010} see Supplementary Information \S 1) but tip the balance of streaming drag forces and acoustic radiation forces so that particles get caught in streaming vortex rolls and don’t equilibrate to static positions. \cite{muller2012}
After the particles reach their equilibrium positions, the samples were immediately cured with a UV LED. 

The ends of the composites were coated with conductive silver epoxy (Circuitworks Conductive Pen) and contacted with probes (contact resistance 2 Ohms determined by calibration to metal with known conductance). The resistance of the samples was measured with a multimeter (Fluke 115) and a current-voltage sweep using a source meter (Kiethley 2636A). The conductivity of the composite samples was determined using the resistance of the sample and using the cross-sectional area of the entire composite (0.9 mm$^2$).
The percolation threshold $\phi_c$ is calculated as the average of the highest volume fraction (volume of carbon to total volume) that results in an open circuit and the lowest volume fraction which results in a closed circuit, with error equal to the half difference between the two. Four measurements were taken within 0.5\% of the observed transition in each of four trials; an overlap was observed in only one of the trials, and three more measurements were made to identify the outlier in that trial.

Bending tests were performed by clamping both ends of patterned fiber-elastomer composite samples with 2.5 mm clamps and cementing them in place with silver-filled epoxy (Chemtronics). The clamps were held in place with micrograbbers on sliding tracks and wired to a multimeter. The elastomer composites were then bent around an insulating mandrel of known diameter and the resistance was recorded. The cycle frequency was 0.5 Hz, measuring resistance at the flat and bent states.

\subsection*{Acknowledgements}
This work was supported by the Institute for Collaborative Biotechnologies through contract no. W911NF-09-D-0001 from the U.S. Army Research Office. N.D.D. was supported by the Institute for Collaborative Biotechnologies through Grant W911NF-09-0001. The work used the Microfluidics Laboratory at the California Nanosystems Institute. The authors acknowledge Trevor Lancon and S. Ali Shojaee at Thermo Fisher for providing and analyzing CT scans, and Chris Bates and Craig Hawker for fruitful discussions. The research reported here made use of shared facilities of the UCSB MRSEC (NSF DMR 1720256), a member of the Materials Research Facilities Network.

\bibliography{MyLibrary6-19} 

\providecommand*{\mcitethebibliography}{\thebibliography}
\csname @ifundefined\endcsname{endmcitethebibliography}
{\let\endmcitethebibliography\endthebibliography}{}
\begin{mcitethebibliography}{52}
\providecommand*{\natexlab}[1]{#1}
\providecommand*{\mciteSetBstSublistMode}[1]{}
\providecommand*{\mciteSetBstMaxWidthForm}[2]{}
\providecommand*{\mciteBstWouldAddEndPuncttrue}
  {\def\EndOfBibitem{\unskip.}}
\providecommand*{\mciteBstWouldAddEndPunctfalse}
  {\let\EndOfBibitem\relax}
\providecommand*{\mciteSetBstMidEndSepPunct}[3]{}
\providecommand*{\mciteSetBstSublistLabelBeginEnd}[3]{}
\providecommand*{\EndOfBibitem}{}
\mciteSetBstSublistMode{f}
\mciteSetBstMaxWidthForm{subitem}
{\alph{mcitesubitemcount})}
\mciteSetBstSublistLabelBeginEnd{\mcitemaxwidthsubitemform\space}
{\relax}{\relax}

\bibitem[Sekitani and Someya(2010)]{sekitani2010}
T.~Sekitani, T.~Someya, \emph{Advanced Materials} \textbf{2010}, \emph{22},
  2228--2246\relax
\mciteBstWouldAddEndPuncttrue
\mciteSetBstMidEndSepPunct{\mcitedefaultmidpunct}
{\mcitedefaultendpunct}{\mcitedefaultseppunct}\relax
\EndOfBibitem
\bibitem[Rogers \emph{et~al.}(2010)Rogers, Someya, and Huang]{rogers2010}
J.~A. Rogers, T.~Someya, Y.~Huang, \emph{Science} \textbf{2010}, \emph{327},
  1603\relax
\mciteBstWouldAddEndPuncttrue
\mciteSetBstMidEndSepPunct{\mcitedefaultmidpunct}
{\mcitedefaultendpunct}{\mcitedefaultseppunct}\relax
\EndOfBibitem
\bibitem[Ray \emph{et~al.}(2019)Ray, Choi, Bandodkar, Krishnan, Gutruf, Tian,
  Ghaffari, and Rogers]{ray2019}
T.~R. Ray, J.~Choi, A.~J. Bandodkar, S.~Krishnan, P.~Gutruf, L.~Tian,
  R.~Ghaffari, J.~A. Rogers, \emph{Chemical Reviews} \textbf{2019}, \emph{119},
  5461--5533\relax
\mciteBstWouldAddEndPuncttrue
\mciteSetBstMidEndSepPunct{\mcitedefaultmidpunct}
{\mcitedefaultendpunct}{\mcitedefaultseppunct}\relax
\EndOfBibitem
\bibitem[Mutiso and Winey(2015)]{mutiso2015}
R.~M. Mutiso, K.~I. Winey, \emph{Progress in Polymer Science} \textbf{2015},
  \emph{40}, 63--84\relax
\mciteBstWouldAddEndPuncttrue
\mciteSetBstMidEndSepPunct{\mcitedefaultmidpunct}
{\mcitedefaultendpunct}{\mcitedefaultseppunct}\relax
\EndOfBibitem
\bibitem[Sekitani \emph{et~al.}(2009)Sekitani, Nakajima, Maeda, Fukushima,
  Aida, Hata, and Someya]{sekitani2009}
T.~Sekitani, H.~Nakajima, H.~Maeda, T.~Fukushima, T.~Aida, K.~Hata, T.~Someya,
  \emph{Nature Materials} \textbf{2009}, \emph{8}, 494--499\relax
\mciteBstWouldAddEndPuncttrue
\mciteSetBstMidEndSepPunct{\mcitedefaultmidpunct}
{\mcitedefaultendpunct}{\mcitedefaultseppunct}\relax
\EndOfBibitem
\bibitem[Niu \emph{et~al.}(2007)Niu, Peng, Liu, Wen, and Sheng]{niu2007}
X.~Z. Niu, S.~L. Peng, L.~Y. Liu, W.~J. Wen, P.~Sheng, \emph{Advanced
  Materials} \textbf{2007}, \emph{19}, 2682--2686\relax
\mciteBstWouldAddEndPuncttrue
\mciteSetBstMidEndSepPunct{\mcitedefaultmidpunct}
{\mcitedefaultendpunct}{\mcitedefaultseppunct}\relax
\EndOfBibitem
\bibitem[Kim \emph{et~al.}(2009)Kim, Zhao, Jang, Lee, Kim, Kim, Ahn, Kim, Choi,
  and Hong]{kim2009}
K.~S. Kim, Y.~Zhao, H.~Jang, S.~Y. Lee, J.~M. Kim, K.~S. Kim, J.-H. Ahn,
  P.~Kim, J.-Y. Choi, B.~H. Hong, \emph{Nature} \textbf{2009}, \emph{457},
  706--710\relax
\mciteBstWouldAddEndPuncttrue
\mciteSetBstMidEndSepPunct{\mcitedefaultmidpunct}
{\mcitedefaultendpunct}{\mcitedefaultseppunct}\relax
\EndOfBibitem
\bibitem[Merilampi \emph{et~al.}(2010)Merilampi, Bj{\"o}rninen, Haukka,
  Ruuskanen, Ukkonen, and Syd{\"a}nheimo]{merilampi2010}
S.~Merilampi, T.~Bj{\"o}rninen, V.~Haukka, P.~Ruuskanen, L.~Ukkonen,
  L.~Syd{\"a}nheimo, \emph{Microelectronics Reliability} \textbf{2010},
  \emph{50}, 2001--2011\relax
\mciteBstWouldAddEndPuncttrue
\mciteSetBstMidEndSepPunct{\mcitedefaultmidpunct}
{\mcitedefaultendpunct}{\mcitedefaultseppunct}\relax
\EndOfBibitem
\bibitem[{de Leon} \emph{et~al.}(2018){de Leon}, Rodier, Bajamundi, Espera,
  Wei, Kwon, Williams, Ilijasic, Advincula, and Pentzer]{deleon2018}
A.~C. {de Leon}, B.~J. Rodier, C.~Bajamundi, A.~Espera, P.~Wei, J.~G. Kwon,
  J.~Williams, F.~Ilijasic, R.~C. Advincula, E.~Pentzer, \emph{ACS Applied
  Energy Materials} \textbf{2018}, \emph{1}, 1726--1733\relax
\mciteBstWouldAddEndPuncttrue
\mciteSetBstMidEndSepPunct{\mcitedefaultmidpunct}
{\mcitedefaultendpunct}{\mcitedefaultseppunct}\relax
\EndOfBibitem
\bibitem[Postiglione \emph{et~al.}(2015)Postiglione, Natale, Griffini, Levi,
  and Turri]{postiglione2015}
G.~Postiglione, G.~Natale, G.~Griffini, M.~Levi, S.~Turri, \emph{Composites
  Part A: Applied Science and Manufacturing} \textbf{2015}, \emph{76},
  110--114\relax
\mciteBstWouldAddEndPuncttrue
\mciteSetBstMidEndSepPunct{\mcitedefaultmidpunct}
{\mcitedefaultendpunct}{\mcitedefaultseppunct}\relax
\EndOfBibitem
\bibitem[Ahn \emph{et~al.}(2009)Ahn, Duoss, Motala, Guo, Park, Xiong, Yoon,
  Nuzzo, Rogers, and Lewis]{ahn2009}
B.~Y. Ahn, E.~B. Duoss, M.~J. Motala, X.~Guo, S.-I. Park, Y.~Xiong, J.~Yoon,
  R.~G. Nuzzo, J.~A. Rogers, J.~A. Lewis, \emph{Science} \textbf{2009},
  \emph{323}, 1590--1593\relax
\mciteBstWouldAddEndPuncttrue
\mciteSetBstMidEndSepPunct{\mcitedefaultmidpunct}
{\mcitedefaultendpunct}{\mcitedefaultseppunct}\relax
\EndOfBibitem
\bibitem[Perelaer \emph{et~al.}(2010)Perelaer, Smith, Mager, Soltman, Volkman,
  Subramanian, Korvink, and Schubert]{perelaer2010}
J.~Perelaer, P.~J. Smith, D.~Mager, D.~Soltman, S.~K. Volkman, V.~Subramanian,
  J.~G. Korvink, U.~S. Schubert, \emph{Journal of Materials Chemistry}
  \textbf{2010}, \emph{20}, 8446\relax
\mciteBstWouldAddEndPuncttrue
\mciteSetBstMidEndSepPunct{\mcitedefaultmidpunct}
{\mcitedefaultendpunct}{\mcitedefaultseppunct}\relax
\EndOfBibitem
\bibitem[Perelaer \emph{et~al.}(2006)Perelaer, {de Gans}, and
  Schubert]{perelaer2006}
J.~Perelaer, B.-J. {de Gans}, U.~S. Schubert, \emph{Advanced Materials}
  \textbf{2006}, \emph{18}, 2101--2104\relax
\mciteBstWouldAddEndPuncttrue
\mciteSetBstMidEndSepPunct{\mcitedefaultmidpunct}
{\mcitedefaultendpunct}{\mcitedefaultseppunct}\relax
\EndOfBibitem
\bibitem[Truby and Lewis(2016)]{truby2016}
R.~L. Truby, J.~A. Lewis, \emph{Nature} \textbf{2016}, \emph{540},
  371--378\relax
\mciteBstWouldAddEndPuncttrue
\mciteSetBstMidEndSepPunct{\mcitedefaultmidpunct}
{\mcitedefaultendpunct}{\mcitedefaultseppunct}\relax
\EndOfBibitem
\bibitem[Dickey(2017)]{dickey2017}
M.~D. Dickey, \emph{Advanced Materials} \textbf{2017}, \emph{29}, 1606425\relax
\mciteBstWouldAddEndPuncttrue
\mciteSetBstMidEndSepPunct{\mcitedefaultmidpunct}
{\mcitedefaultendpunct}{\mcitedefaultseppunct}\relax
\EndOfBibitem
\bibitem[Saengchairat \emph{et~al.}(2017)Saengchairat, Tran, and
  Chua]{saengchairat2017}
N.~Saengchairat, T.~Tran, C.-K. Chua, \emph{Virtual and Physical Prototyping}
  \textbf{2017}, \emph{12}, 31--46\relax
\mciteBstWouldAddEndPuncttrue
\mciteSetBstMidEndSepPunct{\mcitedefaultmidpunct}
{\mcitedefaultendpunct}{\mcitedefaultseppunct}\relax
\EndOfBibitem
\bibitem[MacDonald and Wicker(2016)]{macdonald2016}
E.~MacDonald, R.~Wicker, \emph{Science} \textbf{2016}, \emph{353},
  aaf2093--aaf2093\relax
\mciteBstWouldAddEndPuncttrue
\mciteSetBstMidEndSepPunct{\mcitedefaultmidpunct}
{\mcitedefaultendpunct}{\mcitedefaultseppunct}\relax
\EndOfBibitem
\bibitem[Rus and Tolley(2015)]{rus2015}
D.~Rus, M.~T. Tolley, \emph{Nature} \textbf{2015}, \emph{521}, 467--475\relax
\mciteBstWouldAddEndPuncttrue
\mciteSetBstMidEndSepPunct{\mcitedefaultmidpunct}
{\mcitedefaultendpunct}{\mcitedefaultseppunct}\relax
\EndOfBibitem
\bibitem[Valentine \emph{et~al.}(2017)Valentine, Busbee, Boley, Raney, Chortos,
  Kotikian, Berrigan, Durstock, and Lewis]{valentine2017}
A.~D. Valentine, T.~A. Busbee, J.~W. Boley, J.~R. Raney, A.~Chortos,
  A.~Kotikian, J.~D. Berrigan, M.~F. Durstock, J.~A. Lewis, \emph{Advanced
  Materials} \textbf{2017}, \emph{29}, 1703817\relax
\mciteBstWouldAddEndPuncttrue
\mciteSetBstMidEndSepPunct{\mcitedefaultmidpunct}
{\mcitedefaultendpunct}{\mcitedefaultseppunct}\relax
\EndOfBibitem
\bibitem[Kokkinis \emph{et~al.}(2015)Kokkinis, Schaffner, and
  Studart]{kokkinis2015}
D.~Kokkinis, M.~Schaffner, A.~R. Studart, \emph{Nature Communications}
  \textbf{2015}, \emph{6}, year\relax
\mciteBstWouldAddEndPuncttrue
\mciteSetBstMidEndSepPunct{\mcitedefaultmidpunct}
{\mcitedefaultendpunct}{\mcitedefaultseppunct}\relax
\EndOfBibitem
\bibitem[Martin \emph{et~al.}(2015)Martin, Fiore, and Erb]{Martin2015}
J.~J. Martin, B.~E. Fiore, R.~M. Erb, \emph{Nature Communications}
  \textbf{2015}, \emph{6}, year\relax
\mciteBstWouldAddEndPuncttrue
\mciteSetBstMidEndSepPunct{\mcitedefaultmidpunct}
{\mcitedefaultendpunct}{\mcitedefaultseppunct}\relax
\EndOfBibitem
\bibitem[Loth \emph{et~al.}(2012)Loth, Baumann, Lutz, Eigler, and
  Heinrich]{loth2012}
E.~Loth, S.~Baumann, C.~P. Lutz, D.~M. Eigler, S.~Heinrich, \emph{Science}
  \textbf{2012}, \emph{335}, 196--199\relax
\mciteBstWouldAddEndPuncttrue
\mciteSetBstMidEndSepPunct{\mcitedefaultmidpunct}
{\mcitedefaultendpunct}{\mcitedefaultseppunct}\relax
\EndOfBibitem
\bibitem[Compton and Lewis(2014)]{compton2014}
B.~G. Compton, J.~A. Lewis, \emph{Advanced Materials} \textbf{2014}, \emph{26},
  5930--5935\relax
\mciteBstWouldAddEndPuncttrue
\mciteSetBstMidEndSepPunct{\mcitedefaultmidpunct}
{\mcitedefaultendpunct}{\mcitedefaultseppunct}\relax
\EndOfBibitem
\bibitem[Tekinalp \emph{et~al.}(2014)Tekinalp, Kunc, {Velez-Garcia}, Duty,
  Love, Naskar, Blue, and Ozcan]{tekinalp2014}
H.~L. Tekinalp, V.~Kunc, G.~M. {Velez-Garcia}, C.~E. Duty, L.~J. Love, A.~K.
  Naskar, C.~A. Blue, S.~Ozcan, \emph{Composites Science and Technology}
  \textbf{2014}, \emph{105}, 144--150\relax
\mciteBstWouldAddEndPuncttrue
\mciteSetBstMidEndSepPunct{\mcitedefaultmidpunct}
{\mcitedefaultendpunct}{\mcitedefaultseppunct}\relax
\EndOfBibitem
\bibitem[Lewicki \emph{et~al.}(2017)Lewicki, Rodriguez, Zhu, Worsley, Wu,
  Kanarska, Horn, Duoss, Ortega, Elmer, Hensleigh, Fellini, and
  King]{lewicki2017}
J.~P. Lewicki, J.~N. Rodriguez, C.~Zhu, M.~A. Worsley, A.~S. Wu, Y.~Kanarska,
  J.~D. Horn, E.~B. Duoss, J.~M. Ortega, W.~Elmer, R.~Hensleigh, R.~A. Fellini,
  M.~J. King, \emph{Scientific Reports} \textbf{2017}, \emph{7}, year\relax
\mciteBstWouldAddEndPuncttrue
\mciteSetBstMidEndSepPunct{\mcitedefaultmidpunct}
{\mcitedefaultendpunct}{\mcitedefaultseppunct}\relax
\EndOfBibitem
\bibitem[Xiong \emph{et~al.}(2016)Xiong, Liu, Jiang, Zhou, Li, Jiang, Silvain,
  and Lu]{xiong2016}
W.~Xiong, Y.~Liu, L.~J. Jiang, Y.~S. Zhou, D.~W. Li, L.~Jiang, J.-F. Silvain,
  Y.~F. Lu, \emph{Advanced Materials} \textbf{2016}, \emph{28},
  2002--2009\relax
\mciteBstWouldAddEndPuncttrue
\mciteSetBstMidEndSepPunct{\mcitedefaultmidpunct}
{\mcitedefaultendpunct}{\mcitedefaultseppunct}\relax
\EndOfBibitem
\bibitem[Collino \emph{et~al.}(2018)Collino, Ray, Friedrich, Cornell, Meinhart,
  and Begley]{collino2018}
R.~R. Collino, T.~R. Ray, L.~M. Friedrich, J.~D. Cornell, C.~D. Meinhart, M.~R.
  Begley, \emph{Materials Research Letters} \textbf{2018}, \emph{6},
  191--198\relax
\mciteBstWouldAddEndPuncttrue
\mciteSetBstMidEndSepPunct{\mcitedefaultmidpunct}
{\mcitedefaultendpunct}{\mcitedefaultseppunct}\relax
\EndOfBibitem
\bibitem[Greenhall \emph{et~al.}(2014)Greenhall, Guevara~Vasquez, and
  Raeymaekers]{greenhall2014}
J.~Greenhall, F.~Guevara~Vasquez, B.~Raeymaekers, \emph{Applied Physics
  Letters} \textbf{2014}, \emph{105}, 144105\relax
\mciteBstWouldAddEndPuncttrue
\mciteSetBstMidEndSepPunct{\mcitedefaultmidpunct}
{\mcitedefaultendpunct}{\mcitedefaultseppunct}\relax
\EndOfBibitem
\bibitem[Reyes \emph{et~al.}(2018)Reyes, Fu, Suthanthiraraj, Owens, Shields,
  L{\'o}pez, Charbonneau, and Wiley]{reyes2018}
C.~Reyes, L.~Fu, P.~P.~A. Suthanthiraraj, C.~E. Owens, C.~W. Shields, G.~P.
  L{\'o}pez, P.~Charbonneau, B.~J. Wiley, \emph{Particle \& Particle Systems
  Characterization} \textbf{2018}, \emph{35}, 1700470\relax
\mciteBstWouldAddEndPuncttrue
\mciteSetBstMidEndSepPunct{\mcitedefaultmidpunct}
{\mcitedefaultendpunct}{\mcitedefaultseppunct}\relax
\EndOfBibitem
\bibitem[Hags{\"a}ter \emph{et~al.}(2007)Hags{\"a}ter, Jensen, Bruus, and
  Kutter]{hagsater2007}
S.~M. Hags{\"a}ter, T.~G. Jensen, H.~Bruus, J.~P. Kutter, \emph{Lab on a Chip}
  \textbf{2007}, \emph{7}, 1336\relax
\mciteBstWouldAddEndPuncttrue
\mciteSetBstMidEndSepPunct{\mcitedefaultmidpunct}
{\mcitedefaultendpunct}{\mcitedefaultseppunct}\relax
\EndOfBibitem
\bibitem[Collino \emph{et~al.}(2015)Collino, Ray, Fleming, Sasaki,
  {Haj-Hariri}, and Begley]{collino2015}
R.~R. Collino, T.~R. Ray, R.~C. Fleming, C.~H. Sasaki, H.~{Haj-Hariri}, M.~R.
  Begley, \emph{Extreme Mechanics Letters} \textbf{2015}, \emph{5},
  37--46\relax
\mciteBstWouldAddEndPuncttrue
\mciteSetBstMidEndSepPunct{\mcitedefaultmidpunct}
{\mcitedefaultendpunct}{\mcitedefaultseppunct}\relax
\EndOfBibitem
\bibitem[Collino \emph{et~al.}(2016)Collino, Ray, Fleming, Cornell, Compton,
  and Begley]{collino2016}
R.~R. Collino, T.~R. Ray, R.~C. Fleming, J.~D. Cornell, B.~G. Compton, M.~R.
  Begley, \emph{Extreme Mechanics Letters} \textbf{2016}, \emph{8},
  96--106\relax
\mciteBstWouldAddEndPuncttrue
\mciteSetBstMidEndSepPunct{\mcitedefaultmidpunct}
{\mcitedefaultendpunct}{\mcitedefaultseppunct}\relax
\EndOfBibitem
\bibitem[Greenhall and Raeymaekers(2017)]{greenhall2017}
J.~Greenhall, B.~Raeymaekers, \emph{Advanced Materials Technologies}
  \textbf{2017}, \emph{2}, 1700122\relax
\mciteBstWouldAddEndPuncttrue
\mciteSetBstMidEndSepPunct{\mcitedefaultmidpunct}
{\mcitedefaultendpunct}{\mcitedefaultseppunct}\relax
\EndOfBibitem
\bibitem[Yunus \emph{et~al.}(2017)Yunus, Sohrabi, He, Shi, and Liu]{yunus2017}
D.~E. Yunus, S.~Sohrabi, R.~He, W.~Shi, Y.~Liu, \emph{Journal of Micromechanics
  and Microengineering} \textbf{2017}, \emph{27}, 045016\relax
\mciteBstWouldAddEndPuncttrue
\mciteSetBstMidEndSepPunct{\mcitedefaultmidpunct}
{\mcitedefaultendpunct}{\mcitedefaultseppunct}\relax
\EndOfBibitem
\bibitem[Greenhall \emph{et~al.}(2016)Greenhall, Guevara~Vasquez, and
  Raeymaekers]{greenhall2016}
J.~Greenhall, F.~Guevara~Vasquez, B.~Raeymaekers, \emph{Applied Physics
  Letters} \textbf{2016}, \emph{108}, 103103\relax
\mciteBstWouldAddEndPuncttrue
\mciteSetBstMidEndSepPunct{\mcitedefaultmidpunct}
{\mcitedefaultendpunct}{\mcitedefaultseppunct}\relax
\EndOfBibitem
\bibitem[Melde \emph{et~al.}(2016)Melde, Mark, Qiu, and Fischer]{melde2016}
K.~Melde, A.~G. Mark, T.~Qiu, P.~Fischer, \emph{Nature} \textbf{2016},
  \emph{537}, 518--522\relax
\mciteBstWouldAddEndPuncttrue
\mciteSetBstMidEndSepPunct{\mcitedefaultmidpunct}
{\mcitedefaultendpunct}{\mcitedefaultseppunct}\relax
\EndOfBibitem
\bibitem[Friedrich \emph{et~al.}(2017)Friedrich, Collino, Ray, and
  Begley]{friedrich2017}
L.~Friedrich, R.~Collino, T.~Ray, M.~Begley, \emph{Sensors and Actuators A:
  Physical} \textbf{2017}, \emph{268}, 213--221\relax
\mciteBstWouldAddEndPuncttrue
\mciteSetBstMidEndSepPunct{\mcitedefaultmidpunct}
{\mcitedefaultendpunct}{\mcitedefaultseppunct}\relax
\EndOfBibitem
\bibitem[Derby(2010)]{derby2010}
B.~Derby, \emph{Annual Review of Materials Research} \textbf{2010}, \emph{40},
  395--414\relax
\mciteBstWouldAddEndPuncttrue
\mciteSetBstMidEndSepPunct{\mcitedefaultmidpunct}
{\mcitedefaultendpunct}{\mcitedefaultseppunct}\relax
\EndOfBibitem
\bibitem[Shofner \emph{et~al.}(2003)Shofner, Lozano,
  {Rodr{\'i}guez-Mac{\'i}as}, and Barrera]{shofner2003}
M.~L. Shofner, K.~Lozano, F.~J. {Rodr{\'i}guez-Mac{\'i}as}, E.~V. Barrera,
  \emph{Journal of Applied Polymer Science} \textbf{2003}, \emph{89},
  3081--3090\relax
\mciteBstWouldAddEndPuncttrue
\mciteSetBstMidEndSepPunct{\mcitedefaultmidpunct}
{\mcitedefaultendpunct}{\mcitedefaultseppunct}\relax
\EndOfBibitem
\bibitem[Ning \emph{et~al.}(2015)Ning, Cong, Qiu, Wei, and Wang]{ning2015}
F.~Ning, W.~Cong, J.~Qiu, J.~Wei, S.~Wang, \emph{Composites Part B:
  Engineering} \textbf{2015}, \emph{80}, 369--378\relax
\mciteBstWouldAddEndPuncttrue
\mciteSetBstMidEndSepPunct{\mcitedefaultmidpunct}
{\mcitedefaultendpunct}{\mcitedefaultseppunct}\relax
\EndOfBibitem
\bibitem[King(1934)]{king1934}
L.~V. King, \emph{Proceedings of the Royal Society A: Mathematical Physical and
  Engineering Sciences} \textbf{1934}, \emph{147}, 212--240\relax
\mciteBstWouldAddEndPuncttrue
\mciteSetBstMidEndSepPunct{\mcitedefaultmidpunct}
{\mcitedefaultendpunct}{\mcitedefaultseppunct}\relax
\EndOfBibitem
\bibitem[GOR'KOV(1962)]{20000152384}
L.~P. GOR'KOV, \emph{Sov. Phys. Dokl.} \textbf{1962}, \emph{6}, 773--775\relax
\mciteBstWouldAddEndPuncttrue
\mciteSetBstMidEndSepPunct{\mcitedefaultmidpunct}
{\mcitedefaultendpunct}{\mcitedefaultseppunct}\relax
\EndOfBibitem
\bibitem[Bruus(2012)]{bruus2012}
H.~Bruus, \emph{Lab on a Chip} \textbf{2012}, \emph{12}, 1014\relax
\mciteBstWouldAddEndPuncttrue
\mciteSetBstMidEndSepPunct{\mcitedefaultmidpunct}
{\mcitedefaultendpunct}{\mcitedefaultseppunct}\relax
\EndOfBibitem
\bibitem[Foygel \emph{et~al.}(2005)Foygel, Morris, Anez, French, and
  Sobolev]{foygel2005}
M.~Foygel, R.~D. Morris, D.~Anez, S.~French, V.~L. Sobolev, \emph{Physical
  Review B} \textbf{2005}, \emph{71}, year\relax
\mciteBstWouldAddEndPuncttrue
\mciteSetBstMidEndSepPunct{\mcitedefaultmidpunct}
{\mcitedefaultendpunct}{\mcitedefaultseppunct}\relax
\EndOfBibitem
\bibitem[Yamahira \emph{et~al.}(2000)Yamahira, Hatanaka, Kuwabara, and
  Asai]{yamahira2000}
S.~Yamahira, S.-i. Hatanaka, M.~Kuwabara, S.~Asai, \emph{Japanese Journal of
  Applied Physics} \textbf{2000}, \emph{39}, 3683--3687\relax
\mciteBstWouldAddEndPuncttrue
\mciteSetBstMidEndSepPunct{\mcitedefaultmidpunct}
{\mcitedefaultendpunct}{\mcitedefaultseppunct}\relax
\EndOfBibitem
\bibitem[Bauhofer and Kovacs(2009)]{bauhofer2009}
W.~Bauhofer, J.~Z. Kovacs, \emph{Composites Science and Technology}
  \textbf{2009}, \emph{69}, 1486--1498\relax
\mciteBstWouldAddEndPuncttrue
\mciteSetBstMidEndSepPunct{\mcitedefaultmidpunct}
{\mcitedefaultendpunct}{\mcitedefaultseppunct}\relax
\EndOfBibitem
\bibitem[Levard \emph{et~al.}(2012)Levard, Diglio, Lu, Rahn, and
  Zhang]{levard2012}
T.~Levard, P.~J. Diglio, S.-G. Lu, C.~D. Rahn, Q.~M. Zhang, \emph{Smart
  Materials and Structures} \textbf{2012}, \emph{21}, 012001\relax
\mciteBstWouldAddEndPuncttrue
\mciteSetBstMidEndSepPunct{\mcitedefaultmidpunct}
{\mcitedefaultendpunct}{\mcitedefaultseppunct}\relax
\EndOfBibitem
\bibitem[Cort{\'e}s and Moreno(2003)]{cortes2003}
M.~T. Cort{\'e}s, J.~C. Moreno, \emph{e-Polymers} \textbf{2003}, \emph{3},
  year\relax
\mciteBstWouldAddEndPuncttrue
\mciteSetBstMidEndSepPunct{\mcitedefaultmidpunct}
{\mcitedefaultendpunct}{\mcitedefaultseppunct}\relax
\EndOfBibitem
\bibitem[Iranmanesh \emph{et~al.}(2013)Iranmanesh, Barnkob, Bruus, and
  Wiklund]{iranmanesh2013}
I.~Iranmanesh, R.~Barnkob, H.~Bruus, M.~Wiklund, \emph{Journal of
  Micromechanics and Microengineering} \textbf{2013}, \emph{23}, 105002\relax
\mciteBstWouldAddEndPuncttrue
\mciteSetBstMidEndSepPunct{\mcitedefaultmidpunct}
{\mcitedefaultendpunct}{\mcitedefaultseppunct}\relax
\EndOfBibitem
\bibitem[Muller \emph{et~al.}(2012)Muller, Barnkob, Jensen, and
  Bruus]{muller2012}
P.~B. Muller, R.~Barnkob, M.~J.~H. Jensen, H.~Bruus, \emph{Lab on a Chip}
  \textbf{2012}, \emph{12}, 4617\relax
\mciteBstWouldAddEndPuncttrue
\mciteSetBstMidEndSepPunct{\mcitedefaultmidpunct}
{\mcitedefaultendpunct}{\mcitedefaultseppunct}\relax
\EndOfBibitem
\bibitem[Han \emph{et~al.}(2017)Han, Van~Ha, and Jaeger]{han2017}
E.~Han, N.~Van~Ha, H.~M. Jaeger, \emph{Soft Matter} \textbf{2017}, \emph{13},
  3506--3513\relax
\mciteBstWouldAddEndPuncttrue
\mciteSetBstMidEndSepPunct{\mcitedefaultmidpunct}
{\mcitedefaultendpunct}{\mcitedefaultseppunct}\relax
\EndOfBibitem
\bibitem[Barnkob \emph{et~al.}(2010)Barnkob, Augustsson, Laurell, and
  Bruus]{barnkob2010}
R.~Barnkob, P.~Augustsson, T.~Laurell, H.~Bruus, \emph{Lab on a Chip}
  \textbf{2010}, \emph{10}, 563\relax
\mciteBstWouldAddEndPuncttrue
\mciteSetBstMidEndSepPunct{\mcitedefaultmidpunct}
{\mcitedefaultendpunct}{\mcitedefaultseppunct}\relax
\EndOfBibitem
\end{mcitethebibliography}
\bibliographystyle{angew}
\end{document}